\documentstyle[epsfig,sprocl]{article}

\def\Journal#1#2#3#4{{#1} {\bf #2}, #3 (#4)}

\def\NPB{{\em Nucl. Phys.} B}
\def\PLB{{\em Phys. Lett.}  B}
\def\PRL{\em Phys. Rev. Lett.}
\def\PRD{{\em Phys. Rev.} D}

\def\be{\begin{equation}}
\def\ee{\end{equation}}
\def\bea{\begin{eqnarray}}
\def\eea{\end{eqnarray}}

\begin{document}

\title{THE ISING MODEL AND BUBBLES IN THE QUARK--GLUON PLASMA\footnote{Talk given
at {\em Strong and Electroweak Matter}, Eger, Hungary, 21--25 May 1997}}

\author{ B. SVETITSKY }

\address{School of Physics and Astronomy, Raymond and Beverly Sackler
Faculty of Exact Sciences,
Tel Aviv University, 69978 Tel Aviv, Israel}

\maketitle\abstracts{
I review evidence for the stability of bubbles in the quark--gluon plasma
near the confinement phase transition.
In analogy with the much-studied oil--water emulsions, this raises the possibility
that there are many phases between the pure plasma and the pure hadron gas,
characterized by spontaneous inhomogeneity and modulation.
In studying emulsions, statistical physicists have 
reproduced many of their phases with
microscopic models based on
Ising-like theories with competing interactions.
Hence we seek an effective Ising Hamiltonian for the SU(3) gauge theory
near its transition.
}
  
\section{Background}
I report here on efforts by Nathan Weiss and myself to construct
an Ising effective Hamiltonian for the SU(3) gauge theory near its confinement
phase transition.\cite{SW}
I will devote most of the talk to presenting the motivation for our
work.
Its origin is in hints of strange doings in the physics of
bubbles in the quark--gluon plasma, hints that raise the possibility of
a wealth of phases in the neighborhood of the confinement
transition.

\subsection{Bubbles in the plasma}

\begin{figure}
\centerline{\epsfig{file=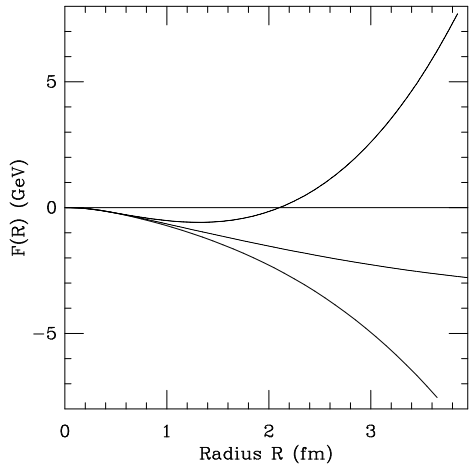,height=2.5in}}
\caption{Free energy of a bubble in the plasma, calculated in the
bag model.
The three curves are, top to bottom, for temperatures just above, at, and just
below the transition.
\label{FR}}
\end{figure}

The bag model provided the first hint of unconventional physics associated
with bubbles in the quark--gluon plasma.
Mardor and Svetitsky\cite{MS} calculated the free energy $F$ of a bubble of
radius $R$ containing a pion gas, surrounded by plasma, at temperatures near 
the transition.
The result, shown in Fig.~\ref{FR}, is that $F(R)$ just above the transition
has a minimum at non-zero $R$.
This means that a bubble, instead of entirely shrinking away above the 
transition, will shrink to a fixed radius and stay there;
moreover, each bubble lowers the total free energy of the system, so
that many bubbles will form spontaneously, limited only by the (unknown)
bubble--bubble interaction energy.
The pure plasma is thus unstable against the formation of a {\em Swiss
cheese}.\cite{Daze}

Looking a bit more deeply, Mardor and Svetitsky considered an expansion
in inverse powers of $R$, beginning with volume, surface, and ``curvature''
terms,
\begin{equation}
F=\Delta P\cdot\frac43\pi R^3+\sigma\cdot4\pi R^2+\alpha\cdot8\pi R+\cdots\ .
\label{F1}
\end{equation}
A local version of Eq.~\ref{F1} applies to an interface of arbitrary shape:
\begin{equation}
F=\Delta P\,V+\sigma\int dS+\alpha\int dS\,\left(\frac1{R_1}+\frac1{R_2}\right)
+\cdots\ ,
\label{Flocal}
\end{equation}
where the last term shown is an integral of the extrinsic curvature,
specified by the local principal radii.
The bag result is that (in either equation) $\sigma$ is small (compared to
the natural scale $B^{3/4}$) while $\alpha$ is not, and furthermore that
$\alpha<0$ for hadron bubbles in the plasma.\footnote{$\alpha$ 
flips sign and becomes positive for plasma droplets in the
hadron gas.}

All this could be dismissed as an oddity of the bag model until lattice
calculations were done.
So far only the pure gauge theory has been studied, but indeed the
surface tension\cite{sigma} $\sigma$ turns out to be perhaps two orders of magnitude
smaller than expected ($\sigma\simeq0.02\, T_c^3$) and a study of 
actual spherical bubbles\cite{sigma} 
showed a negative curvature coefficient $\alpha$ and
perhaps even a non-trivial minimum in $F(R)$.
It is worth noting that the bubble interface on the lattice is quite thick,
so these lattice calculations go beyond the handwaving associated with the
thin wall of the bag model.

\subsection{Oil--water emulsions}

Instead of a mixture of the low- and high-temperature phases of QCD,
consider a more down-to-earth mixture of water and oil.\cite{Safran}
The relative amounts can be controlled via the two fluids' chemical
potentials; the interface area between the two can be controlled by
adding soap, with its own chemical potential.
The interface acquires a spontaneous curvature, which can be adjusted
by adding salt to the water.

One approach to the study of such mixtures is to write the free energy
of a single interface as an expansion of the form of Eq.~\ref{Flocal}
(with additional terms, quadratic in the curvature, to ensure stability).
A large variety of phases has been demonstrated in this way, including
global separation of the oil and water; spherical bubbles; cylindrical
bubbles; planar lamellae; and interpenetrating percolation networks.
One may also write down {\em local\/} models for these systems,\cite{Widom}
typically by defining a spin variable $\sigma$ which is $\pm1$ for oil and
water.
An effective Hamiltonian might be
\begin{equation}
H_{\rm eff}=J\sum_{\langle ij\rangle}\sigma_i\sigma_j
+h\sum\sigma_i+\gamma\sum_{{\rm nnn}}\sigma_i\sigma_k
+\delta\sum\sigma_i\sigma_j\sigma_k\ ,
\label{Heff1}
\end{equation}
with a negative ({\em i.e.,} ferromagnetic) nearest-neighbor coupling $J$ and
a {\em competing,} positive (antiferromagnetic) next-nearest-neighbor
coupling $\gamma$.
The competition between couplings leads to long-range domain structure,
and the odd terms contribute spontaneous curvature to the interfaces.
If we can derive an effective Hamiltonian of this form for
the SU(3) gauge theory, we can see whether it possesses the competing
interactions needed to establish domain structure in equilibrium.

\section{Effective Hamiltonian for SU(3) gauge theory}

\begin{figure}
\centerline{\epsfig{figure=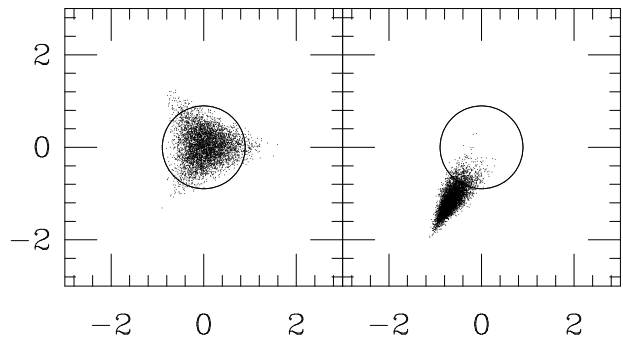,height=2.0in}}
\caption{Distribution of $L({\bf x})$ in the complex plane, after averaging
over blocks of 8 sites, in the confining phase (left) and in the non-confining
phase (right).
Circles have radius $r_\sigma^2=0.8$.
\label{Lfig}}
\end{figure}

We map configurations of the $d=4$ SU(3) gauge theory to Ising spins
in 3 dimensions $\sigma=\pm1$ as follows.
We identify confining and non-confining domains according to the local
value of the Wilson line,
\begin{equation}
L({\bf x})={\rm Tr}\,{\rm P}\,\exp i\int_0^\beta dt\, A_0({\bf x},t)\ ,
\end{equation}
with $\sigma({\bf x})=+1$ if $|L({\bf x})|>r_\sigma$ (with $r_\sigma$ suitably 
chosen) and $\sigma=-1$ otherwise.
In order to obtain clear separation between confining and non-confining
domains, we smear $L$ over a $2\times2\times2$ block before mapping it to
$\sigma$, as shown in Fig.~\ref{Lfig}.
Configurations of $L$, generated by Monte Carlo simulation of the SU(3)
gauge theory, thus yield configurations of $\sigma$; these in turn yield
an effective Hamiltonian $H_{\rm eff}[\sigma]$ along the lines of 
Eq.~\ref{Heff1}
via solution of the lattice Schwinger-Dyson equations.\cite{SD}

Coefficients in $H_{\rm eff}[\sigma]$ are shown in the table.
The first two lines are the couplings for the theory at the phase transition,
$\beta=5.091$ on our lattice, when it is approached from the ordered and
disordered sides, respectively.
$L$ has been smeared, as noted, and a $2\times16\times16\times16$ lattice
gauge theory has been mapped to a $16\times16\times16$ Ising theory.
We note competition between the ferromagnetic nn coupling and the
antiferromagnetic $4^{\rm th}$-neighbor (two-link) coupling, but the latter
is uncomfortably strong, indicating that perhaps more couplings should
be retained.\footnote{In fact the Ising model with these couplings gives a
fairly poor match to the expectation values in the gauge theory.}
The next two lines in the table result upon decimating the spins, 
giving an $8\times8\times8$ Ising theory.
The couplings are short-ranged, but competition has disappeared.

\begin{table}
\label{table1}
\caption{Couplings for the effective Ising Hamiltonian.
The first two lines reflect smearing of $L$; the last two lines reflect smearing
and decimation.}
\vskip\baselineskip
\begin{center}
\begin{tabular}{|cccccccc|}
\hline
&&&&&&&\\
&$h$&nn&nnn&{$3^{\rm rd}$}&{$4^{\rm th}$}&3-spin$_1$&3-spin$_2$\\
\hline
{ordered:}&-0.05&-0.46&-0.05&0.04&0.15&-0.006&0.006\\
{disordered:}&0.14&-0.39&-0.03&0.05&0.13&0.02&0.02\\
\hline
{ordered:}&0.03&-0.13&-0.02&-0.006&0.002&0.004&-0.003\\
{disordered:}&0&-0.24&-0.05&0&0&-0.02&0\\
\hline
\end{tabular}
\end{center}
\end{table}
In both cases the Ising couplings are discontinuous as we
cross the phase transition.
This is inconsistent with the expectation that we should obtain a single
Ising theory whose first-order phase boundary produces the confinement
phase transition.
It is also inconsistent with a theorem proved by van Enter, Fern\'andez,
and Sokal,\cite{Sokal}
which states (in brief) that effective Hamiltonians are either
continuous or nonsense.
It is possible that $H_{\rm eff}[\sigma]$ may be made continuous
by adding (many) more couplings to it; it is also possible that there
exists {\em no} effective Ising Hamiltonian with sensible interactions.
If the latter should prove true, a possible solution lies in defining
a more complex effective spin, perhaps incorporating a $Z(3)$ degree
of freedom in order to preserve the symmetry of the original gauge
theory.\cite{FOU}

\section*{Acknowledgments}
This work was supported by the Israel Science Foundation under 
Grant \break No.~255/96--1.
Further support was provided by the Basic Research Fund of
Tel Aviv University.


\section*{References}
\begin{thebibliography}{99}
\bibitem{SW}B. Svetitsky and N. Weiss, Tel Aviv preprint TAUP-2421-97 
(May 1997), hep-lat/9705007, to appear in {\em Phys. Rev. D}.
\bibitem{MS}I. Mardor and B. Svetitsky,  \Journal{\PRD}{44}{878}{1991}.
\bibitem{Daze}B. Svetitsky in {\em Hot Summer Daze}, ed. A. Gocksch and R.
Pisarski (World Scientific, Singapore, 1992), p. 204; 
G.~Lana and B.~Svetitsky, \Journal{\PLB}{285}{251}{1992}.
\bibitem{sigma}See references in Ref.~1.
\bibitem{Safran}S. A. Safran, {\em Statistical Thermodynamics of Surfaces,
Interfaces, and Membranes} (Addison-Wesley, Reading, Mass., 1994).
\bibitem{Widom}B. Widom, \Journal{\em J. Chem. Phys.}{84}{6943}{1986};
G. Gompper and M. Schick, \Journal{{\em Phys. Rev.} B}{41}{9148}{1990}.
\bibitem{SD}M.~Falcioni {\em et al.}, 
\Journal{\NPB}{265 [FS15]}{187}{1986};
A.~Gonz\'alez-Arroyo and M.~Okawa, 
\Journal{\PRD}{35}{672}{1987}.

\bibitem{Sokal}A. C. D. van Enter, R. Fern\'andez, and A. D. Sokal,
\Journal{\em J. Stat. Phys.}{72}{879}{1993}.
\bibitem{FOU}M.~Fukugita, M.~Okawa, and A.~Ukawa, \Journal{\PRL}{63}{1768}{1989}.

\end{thebibliography}
\end{document}